\documentstyle[11pt]{article}
\begin{document}

\title{Reply to {\tt cond-mat/0503325}: Comment on `Generating functional analysis of
Minority Games with real market histories'\\ by KH Ho, WC Man, FK
Chow and HF Chau}
\date{}
\author{ACC Coolen\\ Department of Mathematics, King's College London} \maketitle

\noindent  The Comment {\tt cond-mat/0503325} is built around two
core statements, both of which are plainly incorrect. Let me
summarize:
\begin{itemize}
\item
{\em `In his recent paper [1] Coolen investigated the dynamics of
the standard version
    of MG using the real and fake histories by the generating functional method of
    De Dominicis [2]. In particular, he concluded that macroscopic quantities
    for the standard version of MG have the same values for the games using real and
    fake histories in the symmetric phase (that is, when $\alpha<\alpha_c$)
    [1].'}
\\[3mm]
    \underline{Incorrect}: Nowhere in [1] is this either concluded, stated or even hinted at.
\\[3mm]
    The formalism in [1] is first developed for arbitrary $\alpha$ (i.e. for both phases).
    The macroscopic laws are then solved, however, ONLY in the time-translation invariant ergodic regime. No claims,
    or graphs relating to macroscopic observables are presented AT ALL for the non-ergodic
    phase $\alpha<\alpha_c$. The latter phase is only mentioned in footnote 8, which
    states, in contrast with the erroneous claim above, that `below the critical point the
    differences between true and fake history (if any) are confined to dynamical phenomena
    or to states without time-translation invariance'.
\item
    {\em `To arrive at this conclusion he made the bold
assumption that the history correlation time $L_h$ is much less
than the number of players $N$. Consequently, one could well
approximate the finite samples of history occurrence frequencies
by ... (1) ...  While this assumption is valid in the asymmetric
phase (that is, when $\alpha>\alpha_c$), .... eq. (1) does not
hold in the symmetric phase.'}
\\[3mm]
   \underline{Incorrect}: In [1] the relevant assumption (1) was made ONLY for the asymmetric
   phase.
\\[3mm]
    This second claim is once more incomprehensible. The relevant information is not
    exactly hard to find: even the very title of the section of [1] in which assumption (1) appears
    is called `TTI states with short history correlation times', i.e. we are
    dealing with $\alpha>\alpha_c$ since time-translation invariance does not hold for $\alpha<\alpha_c$.
\end{itemize}
It is difficult to see how these mistakes could have been made. A
proper reading of only the discussion section of [1], i.e. one
page of plain text without any mathematical theory, would already
have prevented Ho at al from making the above erroneneous claims
(as there the restriction to TTI solutions is again mentioned
explicitly).

It will be clear that Comments and Replies can be valuable, and
that genuine mistakes in the interpretation of someone's work can
 help to disentangle subtle physical or mathematical arguments.
Unfortunately, the Comment submitted by Ho et al is of a different
type.

\end{document}